# Fewer, Better Frames: A Compute-Normalized Proof of Concept for Coherence-First World-Model Rendering with Model-Guided FSR4 Frame Generation

Independent systems experimentation around Waypoint on a Windows fallback stack using ONNX Runtime + DirectML and an FSR4 DX12 bridge.

**Paweł Katarzyński**
Nicolaus Copernicus University in Toruń, Poland

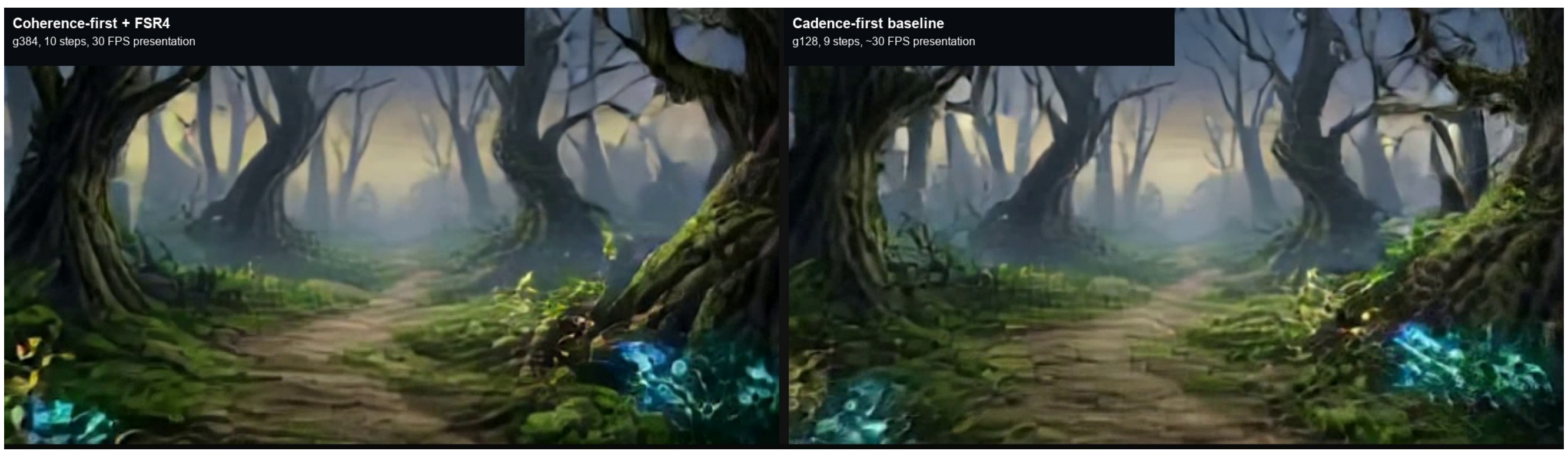

## Abstract

World models are often judged by how much native frame cadence they can generate, but higher nominal frame rate can come at the cost of unstable world states. This article reports a compute-normalized proof of concept implemented using Overworld's Waypoint-1.5 family and WorldEngine runtime [1-4] on a Windows fallback stack with ONNX Runtime + DirectML and an FSR4 DX12 bridge. The tested configuration generates higher-context anchor frames at a 15 FPS presentation-timeline cadence and reconstructs presentation to 30 FPS using latent-delta motion guidance and synthesized depth, then compares that branch against a lower-context cadence-first baseline that generates about 30 FPS natively under matched deterministic rollouts. Across forest, sword, desert, and snow scenes, the coherence-first branch preserves path geometry, object identity, large silhouettes, and depth layering consistently longer than the baseline, which degrades into brightness drift and geometric distortion earlier in the rollout. The paired comparison videos did not reveal a dominant presentation-smoothness artifact large enough to offset that coherence gain. Here compute-normalized means an approximately matched same-GPU, same-timescale operating point for a fixed presentation-duration sequence: one branch generates 15 FPS presentation-timeline anchors, spending roughly twice the generation budget per native frame before FSR4 reconstructs presentation to 30 FPS, while the other uses the same timescale for lower-context native generation at about 30 FPS presentation cadence without frame-generation reconstruction. A secondary, explicitly separate finding is local non-monotonicity: in one sword-scene probe, a heavier `g512` / 12-step configuration underperformed the selected `g384` / 10-step operating point, although both were visually more coherent than the baseline. The g512 run is exploratory and is not part of the main fixed-route comparison. These results support coherence-first allocation as a practical proof-of-concept strategy under limited inference budget, not as a claim of a finished real-time product.

**Reported scope of this evidence**

• A 15 FPS latent-anchor regime reconstructed to a 30 FPS presentation stream. These are presentation-timeline rates, not measured live interactive throughput.

• A Windows proof-of-concept stack using ONNX Runtime + DirectML for world-model inference and an FSR4 DX12 bridge for reconstruction.

• A compute-normalized comparison showing stronger long-horizon coherence than a lower-context cadence-first baseline under the same local time-scaling factor.

Not claimed here

• A production-ready real-time game renderer.

• Native 60 FPS world-model generation or a finished 10 -> 60 pipeline.

• Exact FLOP parity, equal cost per generated native frame, or a complete presentation-stage microbenchmark.

• That the Windows fallback path is the optimized upper bound for Waypoint.

• That the unstable Windows ROCm FP8 attempt was part of the final successful result.

# 1. Introduction

Generative world models promise interactive environments without conventional asset pipelines, but current systems face a practical trade-off: increasing native frame cadence often requires smaller context windows, fewer denoising steps, or both. The result can lose scene stability long before it stops producing frames. Given limited inference budget, a more useful question is what coherence is gained by allocating a fixed same-GPU, same-timescale operating budget to stronger anchor states plus reconstruction rather than lower-context cadence-first generation.

This work studies that question in a concrete implementation based on Overworld's Waypoint-1.5 family and WorldEngine runtime [1-4]. The motivating project vision was roughly 10 -> 60 FPS on AMD hardware. The reported evidence is narrower: a Windows proof-of-concept path that generates 15 FPS latent anchors, reconstructs presentation to 30 FPS, and evaluates the outcome with scripted deterministic rollouts rather than a final live client. The contribution is therefore a systems result about compute allocation, not a claim of a shipping renderer.

In this paper, compute-normalized means that branches are compared on the same GPU, same route, same seed and control script, same target presentation duration, and the same global time-scaling regime on the local hardware. It does not mean exact FLOP parity, equal cost per generated native frame, equal native generation cadence, or measured realtime throughput. The cadence numbers are presentation-timeline rates: the tested GPU generated the evidence offline more slowly than realtime, but the same slowdown factor was applied to both main branches. Under that shared timescale, the coherence-first branch generates half as many native world-model frames and therefore spends roughly twice the generation budget per native frame on context and denoising before reconstructing the missing presentation frames. The cadence-first branch spends the same timescale on the 30 FPS presentation stream natively at lower context and quality. The contribution is not that presentation-stage filtering alone improves outputs. Rather, the pipeline reconstructs presentation from stronger anchor states using motion guidance derived from the model itself.

# 2. Motivation and framing

A native 30-60 FPS number is not meaningful if the world fails to persist from frame to frame. In the paired comparisons, the cadence-first branch does not merely blur; it loses object identity and scene topology. Paths widen or disappear, trunks deform into unstable vertical structures, silhouettes bifurcate, brightness drifts, and foreground/background separation weakens. These are failures of world consistency.

This motivates a coherence-first strategy: spend the limited budget on stronger anchor states, then reconstruct a smoother presentation stream from those anchors. In this setting, reconstruction is not merely a presentation-stage post-process. The world model itself is the source of scene change, so its latent evolution can supply motion cues better aligned with the generated dynamics than purely image-space estimation.

There is also a useful theoretical intuition here, framed as an argument rather than a proof. In a conventional renderer, interpolation artifacts are judged against a privileged missing ground-truth frame. In an autoregressive world model, every displayed frame is already model-mediated. That does not make interpolation artifacts harmless, but it changes the decision rule: the relevant question is whether reconstruction introduces failures more objectionable than the model's own temporal instability. The present comparisons suggest that, at the selected operating point, reconstruction artifacts are not the dominant failure mode.

## 3. Related work

Recent work has rapidly expanded the design space for interactive world models. Genie introduced a foundation-world-model framing in which a model learns action-controllable environments from unlabeled Internet video and a latent action interface [5]. DIAMOND showed that diffusion world models can retain visual detail well enough to improve Atari 100k performance, and also demonstrated an interactive neural game-engine setting on Counter-Strike gameplay [6]. GameNGen pushed the real-time end of the spectrum further by interactively simulating DOOM at over 20 FPS on a single TPU while remaining stable over long trajectories [7]. Oasis and Microsoft's WHAM/WHAMM line further reinforce that learned interactive game/world engines are now a serious systems target rather than a speculative interface [22,23].

More recent frame-centric systems explore a different latency/coherence trade-off. World Labs' RTFM research preview emphasizes real-time, persistent world rendering from a single image and frames the problem explicitly around computational efficiency and persistence [8]. InSpatio-WorldFM likewise adopts a frame-based paradigm, generating frames independently while enforcing spatial consistency with explicit 3D anchors and implicit memory [9]. INSPATIO-WORLD and Infinite-World are especially relevant close neighbors for long-horizon persistence and memory, but they attack the problem through world-model architecture rather than a presentation-reconstruction budget split [24,25].

To position this work against adjacent research, we reviewed four related clusters: interactive world models, sparse/keyframe video interpolation, temporal upscaling and frame generation, and efficient diffusion inference. The closest interpolation-side neighbors include generative keyframe inbetweening, motion-aware latent diffusion interpolation, diffusion video interpolation, arbitrary generative interpolation, flow-guided generative interpolation, and control-guided video diffusion [14-17,26-28]. These works support the general idea that sparse visual anchors and motion/control signals can reconstruct plausible intermediate video, but they do not study a Waypoint-style world-model renderer under matched deterministic rollouts with same-GPU, same-timescale budget allocation. On the graphics side, FSR, DLSS, XeSS, and neural supersampling establish the standard use of motion vectors, depth, and history for temporal reconstruction and frame generation [10,11,18,19,29]. Those systems assume conventional renderer or optical-flow inputs rather than motion guidance derived from a generative world-model state. Efficient diffusion methods such as FlexiDiT, adaptive caching, TeaCache, Pyramid Attention Broadcast, and StreamDiffusion reduce generation cost itself [20,21,30-32], but do not directly compare native world-model cadence against sparse high-quality anchors plus reconstruction.

The present paper is narrower than these works. It does not introduce a new training recipe, a new benchmark, or a new general-purpose world model. Instead, it studies an inference-side systems question inside an existing Waypoint-based stack [1-4]: how much quality is gained by spending a similar same-GPU, same-timescale operating budget on stronger sparse anchors plus reconstruction rather than on lower-context native cadence. On the reconstruction side, AMD's published FSR SDK and FSR4 frame-generation interfaces explicitly use render-resolution motion vectors and depth [10,11], which makes this bridge technically well-posed.

## 4. Prior-art audit summary

A focused citation-snowballing audit was performed around the paper's narrow claim. The audit began from the base implementation sources for Waypoint and WorldEngine [1-4], then expanded through four high-risk neighboring clusters: interactive world models and neural game engines [5-9,22-25], sparse/keyframe interpolation and motion-guided video generation [14-17,26-28], temporal reconstruction and frame-generation APIs [10,11,18,19,29], and efficient diffusion inference / compute-allocation methods [20,21,30-32]. The source registry used during drafting contained 42 entries across these clusters. The purpose was not to prove that no adjacent paper exists, but to test whether the specific claimed bundle had already been reported.

The audit found strong prior art for nearly every component in isolation. World-model papers already show action-controllable generated environments and neural game engines. Interpolation papers already show sparse visual anchors, motion guidance, and plausible in-between video. Graphics reconstruction systems already use motion vectors, depth, history, and frame generation. Efficient diffusion papers already reduce generation cost through caching, attention reuse, or pipeline scheduling. The distinction supported by the audit is narrower: the checked sources did not report a Waypoint-style video world-model renderer that compares lower-context native cadence against fewer stronger generated anchors plus model-state-derived FSR4 frame-generation reconstruction under matched deterministic routes and approximately same-GPU, same-timescale budget.

This means the novelty claim should remain deliberately bounded. The paper does not claim first use of world models, first use of frame interpolation, first use of temporal reconstruction, first use of latent motion guidance, or first efficient diffusion strategy. The supported claim is a systems comparison and interface: using internal world-model state to guide FSR4 reconstruction, then comparing that coherence-first allocation against a cadence-first baseline under the same scripted rollouts. The essential overlap test is included here so the novelty boundary is visible in the main paper itself.

### Table 1. Prior-art direct-overlap test

| Checked cluster | What the prior art covers | Direct-overlap result for this paper |
|---|---|---|
| Interactive world models / neural game engines | Action-controllable generated environments, game-like rollouts, persistence and real-time targets. | Close background, but no checked source reports the sparse-anchor plus FSR4 reconstruction budget comparison. |
| Sparse/keyframe and motion-guided interpolation | Plausible in-between frames from sparse anchors, often with flow, depth, or control guidance. | Strong reconstruction prior, but not a world-model renderer using model-state-derived FSR4 guidance under matched rollouts. |
| FSR / DLSS / XeSS / neural reconstruction | Motion vectors, depth, exposure, masks, history, temporal reconstruction, and frame generation. | Establishes the graphics interface; the checked sources assume conventional renderer or optical-flow inputs rather than world-model latent state. |
| Efficient diffusion inference | Caching, attention reuse, adaptive compute, and streaming pipelines that reduce generation cost itself. | Alternative efficiency route; no checked source directly compares cadence-first native generation against stronger anchors plus reconstruction. |

**Novelty boundary from the audit**

- Claimed: a same-route systems comparison between lower-context native cadence and stronger sparse anchors reconstructed with model-guided FSR4 frame generation.
- Not claimed: invention of world models, frame interpolation, temporal reconstruction, latent diffusion, or efficient diffusion inference.

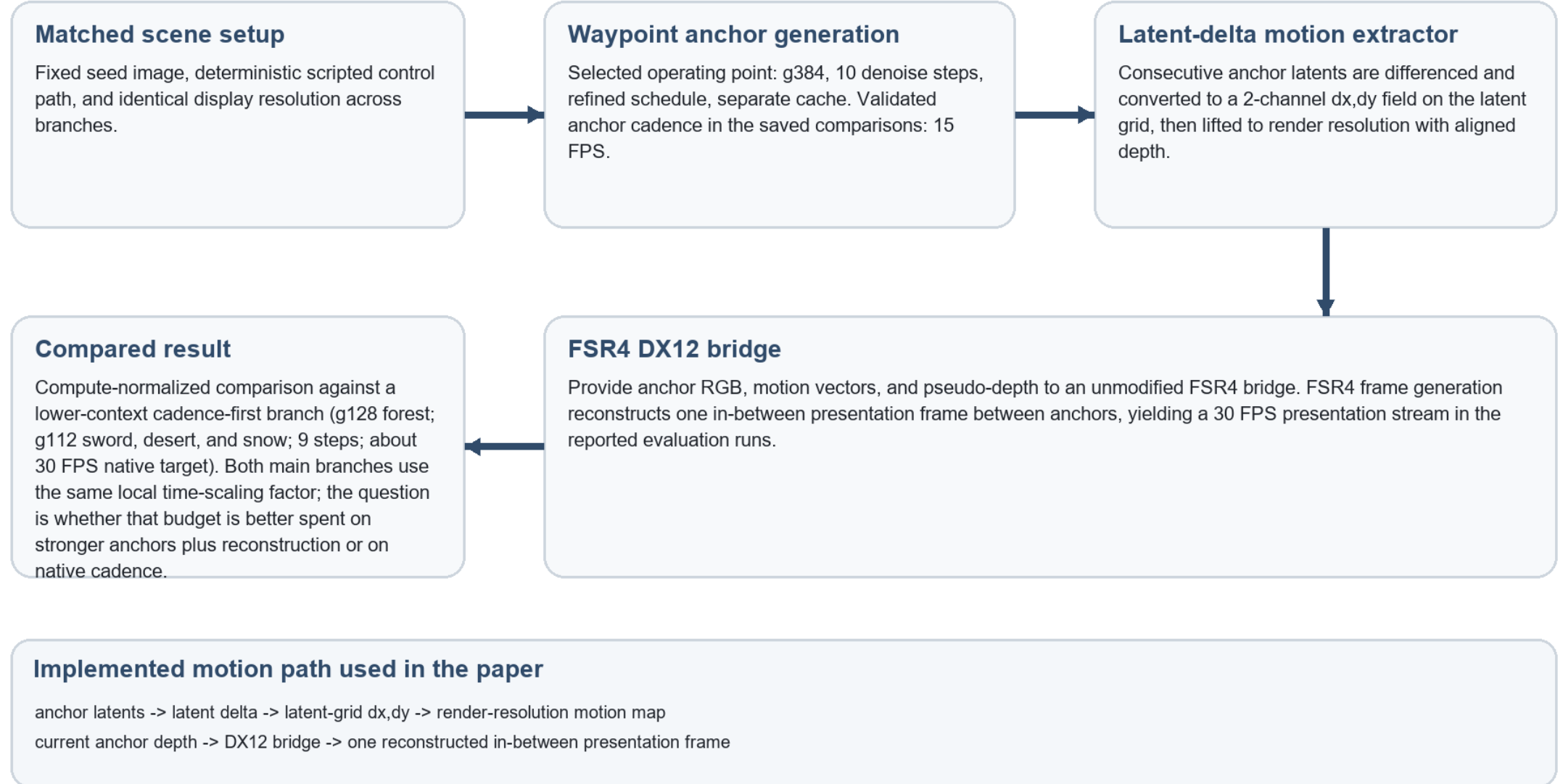

Figure 1. Pipeline overview. The system generates higher-quality anchor frames at a 15 FPS presentation-timeline cadence, converts consecutive anchor latents into a latent-delta motion field, lifts that field to render-resolution motion vectors with aligned synthesized depth, and reconstructs one intermediate presentation frame between anchors using the FSR4 DX12 bridge. FSR4 is unmodified; the experiment supplies model-state-derived motion and pseudo-depth inputs. The contribution is a compute-normalized operating-point strategy and interface, not a claim that the reported Windows stack is the final runtime.

## 5. Pipeline and method

The implemented pipeline couples a Waypoint-based world-model rollout with a reconstruction stage. Each scene begins from a fixed seed image and runs under a deterministic scripted control path so that paired comparisons follow the same trajectory. The selected coherence-first operating point uses `g384` context export, 10 denoise steps, the refined schedule, and separate cache mode. In the reported comparison runs, the model's true latent anchor cadence is 15 FPS on the presentation timeline. One reconstructed frame is generated between successive anchors, producing a 30 FPS presentation stream at 1920 x 1080.

The cadence-first baseline allocates the operating point differently. It uses lower-context 9-step, refined-schedule, separate-cache configurations and generates the approximately 30 FPS presentation stream natively. Its frames are also presented at 1920 x 1080, but it does not use the same coherence-first anchor strategy or FSR4 frame-generation reconstruction. Both main branches are evaluated under the same offline time-scaling regime on the tested GPU. The comparison therefore asks whether a similar same-GPU operating budget is better spent on fewer, better-converged anchor states plus reconstruction or on lower-context cadence-first generation.

### 5.1 Motion-guidance interface

The reported path uses a hand-engineered latent-delta extractor over consecutive world-model anchor latents. Let $z_t$ and $z_{t+1}$ denote consecutive anchor latents on the compressed autoencoder grid, and let $I_t$ and $I_{t+1}$ be their decoded RGB anchors. The motion stage first forms a latent delta field Delta $z_t = z_{t+1} - z_t$. The extractor averages Delta $z_t$ across latent channels, computes finite-difference spatial gradients on that scalar field, and treats the resulting x and y gradients as a 2-channel latent-grid displacement field $d_t$. This extractor was not learned or separately calibrated; it is a deterministic heuristic implemented in the project code.

The field passed to the DX12 bridge is a render-resolution motion map $m_t = -U(8\ d_t)$, where U upsamples the latent-grid displacement to the render grid, 8 is the autoencoder spatial compression factor recorded in the project specification, and the minus sign converts the project's previous-to-current convention to the current-to-previous convention expected by the bridge. Values are clamped to the render-resolution pixel range. An aligned pseudo-depth map $D_t$ is synthesized from the current latent by taking the first latent channel, normalizing it per frame, and upsampling it to render resolution. This depth is a heuristic scene-structure cue, not a measured depth buffer. The bridge then consumes the anchor RGB, motion map, and pseudo-depth to reconstruct one intermediate presentation frame. FSR4 itself is not modified; the experiment supplies model-state-derived motion and pseudo-depth inputs to the bridge. The decoded-frame optical-flow fallback mentioned elsewhere in the project was not used for the reported comparisons.

**Algorithm 1. Implemented motion-guidance path**

- Generate consecutive anchor latents $z_t$ and $z_{t+1}$ and decode anchor RGB frames $I_t$ and $I_{t+1}$.
- Compute Delta $z_t = z_{t+1} - z_t$.
- Apply the latent-delta motion extractor MV(.) to obtain a 2-channel latent-grid field $d_t$.
- Lift $d_t$ to render resolution as $m_t = -U(8\ d_t)$, then clamp to the render-resolution pixel range.
- Synthesize an aligned depth map $D_t$ for the current anchor.
- Provide $(I_t, I_{t+1}, m_t, D_t)$ to the DX12 bridge and reconstruct the in-between presentation frame.

## 6. Experimental setup

All reported comparisons use a Windows fallback stack: ONNX Runtime + DirectML for world-model inference on AMD hardware, plus an FSR4 DX12 bridge for presentation reconstruction. This is a proof-of-concept evaluation path, not the final intended upper-bound runtime for the model. The experiments are deterministic rollouts rather than live interactive sessions.

The main comparison pairs are forest, sword, desert, and snow. Each pair starts from the same seed image, uses the same scripted controls, uses the same ONNX Runtime + DirectML fallback stack, and is presented at 1920 x 1080. In the main quality branch, each rollout consists of 226 anchor frames and 225 inter-anchor motion intervals, yielding 451 output frames. The cadence-first baseline is matched to the same approximate presentation duration and the same local time-scaling factor. Evaluation is primarily qualitative because the central failure mode is long-horizon coherence rather than per-pixel realism.

### Table 2. Compared operating points

| Branch | Context | Steps | Cadence / presentation | Role |
|---|---|---|---|---|
| Coherence-first + FSR4 frame generation | `g384` | `10` | `15 FPS anchors -> 30 FPS presentation` | Selected coherence-first operating point |
| Cadence-first baseline | `g128` forest<br>`g112` sword/desert/snow | `9` | `about 30 FPS native presentation` | Lower-context comparison branch without FSR4 frame generation |
| Heavier probe | `g512` | `12` | `15 FPS anchors -> 30 FPS presentation` | Exploratory only; excluded from main fixed-route comparison |

### Table 3. Hardware and software stack

Driver, operating-system, and runtime version entries are author-recorded environment values. Context, step count, and frame-count values are taken from saved manifests.

| Item | Recorded value |
|---|---|
| GPU | AMD Radeon RX 9070 XT; 16 GB class RDNA 4 card according to AMD product specifications |
| Driver | `32.0.23033.1002` |
| Operating system | Microsoft Windows 11 Pro, build `22631` |
| World-model runtime | ONNX Runtime `1.24.4` with `DmlExecutionProvider` and CPU fallback available |
| Model precision | Exported ONNX models named `worldmodel_fp16`; ROCm FP8 path was not used for the reported assets |
| Presentation bridge | FSR4 DX12 bridge using FSR4 upscaling plus FSR4 frame generation / frame interpolation interfaces through AMD FSR SDK APIs |

### Table 4. Cadence allocation for compute normalization

Compute-normalized means same GPU, same route, same presentation duration, and the same local time-scaling factor for the two main branches. It does not mean equal native generation cadence or measured live interactive throughput: one branch generates 15 FPS anchors with roughly twice the generation budget per native frame and reconstructs to 30 FPS presentation, while the baseline uses the same timescale to generate the approximately 30 FPS presentation stream natively at lower context/quality.

| Branch | Native generation | Presentation | Budget interpretation |
|---|---|---|---|
| Coherence-first + FSR4 frame generation | Generate stronger anchor states at `15 FPS` | FSR4 frame generation reconstructs one in-between frame per anchor interval, yielding `30 FPS presentation` | Spend roughly twice the generation budget per native frame, then reconstruct the missing presentation frames |
| Cadence-first baseline | Generate a lower-context native stream at about `30 FPS presentation` | Present the native stream at about `30 FPS` without FSR4 frame-generation reconstruction | Use the same timescale to generate twice as many lower-context native frames |

## Figure 2. Seed montage

**Seed images used to initialize the deterministic scene rollouts**

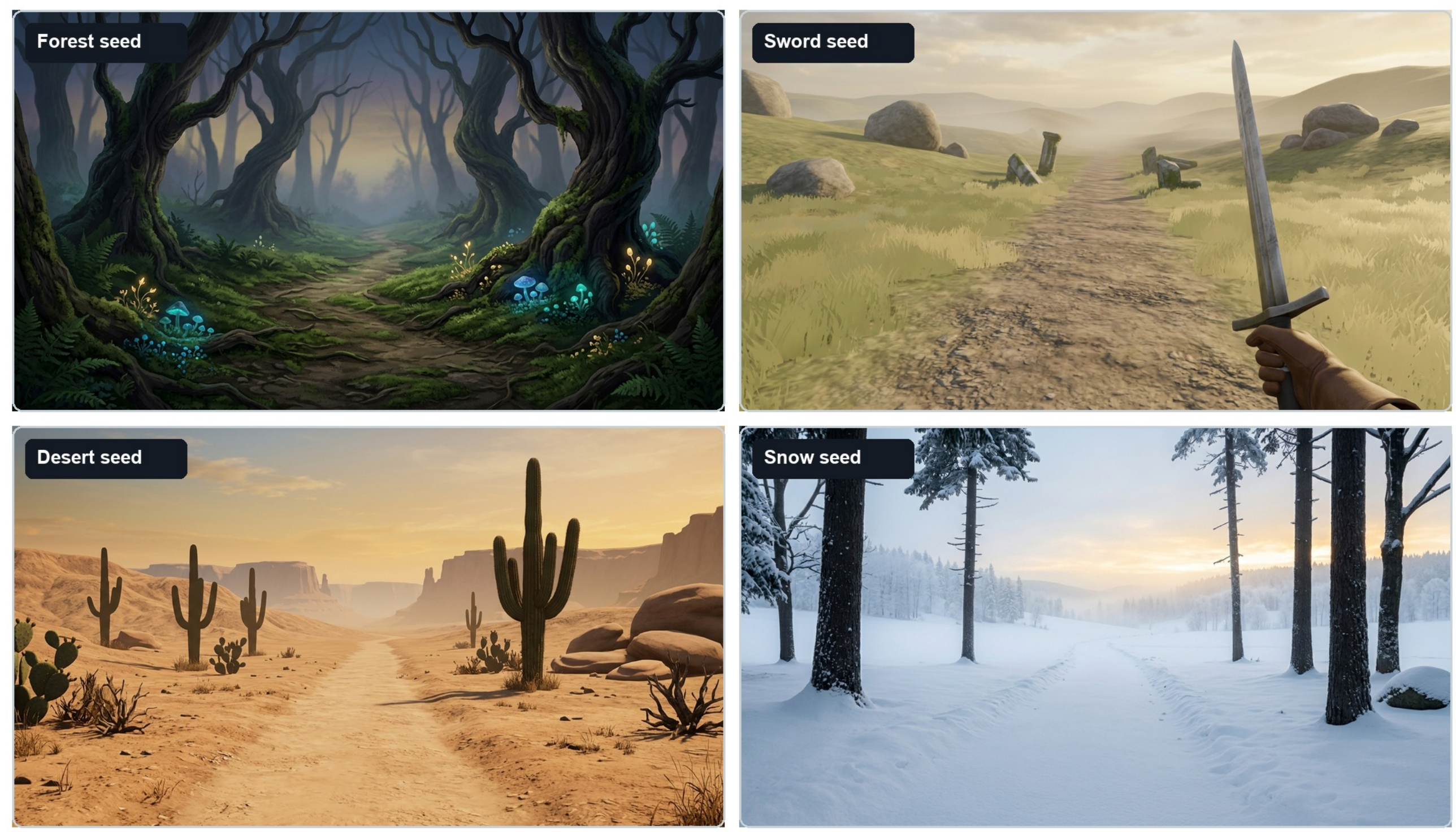

Each paired comparison reuses the same seed image and the same deterministic control script across branches.

Figure 2. Seed images used to initialize the deterministic scene rollouts. Each paired comparison reuses the same seed per scene and the same scripted control path across branches, which matters because the paper's claim is about compute allocation, not about cherry-picked initial states.

### 6.1 Implementation constraints: why this is a fallback result

The original target stack was a native ROCm FP8 runtime on RDNA 4. AMD's Windows ROCm support materials list FP8 as supported on RDNA 4 GPUs while also documenting Windows ROCm limitations [12]. On the tested system, the local ROCm/FP8 path did not yield stable runs suitable for the final paper evidence. The reported evidence therefore comes from an ONNX Runtime + DirectML fallback paired with the FSR4 DX12 bridge.

This constrains how the results should be interpreted. The paper should not be read as showing the upper bound of the hardware or of Waypoint itself. AMD's official RX 9070 XT specifications list native FP8 matrix throughput and dedicated AI hardware [13]. The reported results are therefore better read as a constrained Windows-stack proof of concept: the rendering strategy works on the tested fallback path, even though the originally preferred ROCm path was not stable enough for the final evidence.

#### Implementation constraints note

- The target direction was a native ROCm FP8 runtime on RDNA 4.
- The reported paper evidence comes from a stable ONNX Runtime + DirectML fallback paired with the FSR4 DX12 bridge.
- That should be interpreted as a constrained Windows-stack proof of concept, not as evidence that the hardware lacks the underlying capability.

**Table 5. Scene-by-scene evidence summary**

| Scene | What the coherence-first branch retains | How the baseline fails | Figure role |
|---|---|---|---|
| Forest | Path continuity, trunk identity, fog/depth layering | Warm drift, silhouette deformation, path topology loss | Main figure |
| Sword | Held-item silhouette plus environment persistence | Terrain smear and path loss around a still sword-like anchor | Second main figure |
| Desert | Road shape, cactus identity, mesa silhouette | Object-boundary blur and horizon instability | Supporting figure |
| Snow | Track boundaries, tree lines, scene layering | Blocky reflective artifacts in bright low-texture regions | Supplementary figure |

# 7. Results

Across all tested scenes, the coherence-first branch is visibly more stable than the cadence-first baseline under the matched-route, compute-normalized framing. The gain is not confined to isolated stills. It grows over time. Early frames in many comparisons are reasonably similar between branches, but the cadence-first branch degrades much faster as the rollout continues. This is consistent with the interpretation that the main benefit comes from stronger world-state generation rather than from a one-time presentation-stage filter.

No single failure-time threshold is reported, because the threshold at which a scene loses legibility is scene-dependent and subjective. The supported claim is narrower: under matched rollouts, the cadence-first branch loses scene legibility earlier and more severely, while the coherence-first branch remains readable longer.

## 7.1 Forest: the clearest main result

The forest sequence is the clearest demonstration and serves as the primary figure. Both branches begin in a similar visual regime: a path receding between dark trunks under shallow fog. By the middle of the rollout, the coherence-first branch still preserves path continuity, trunk identity, and layered atmospheric depth. The baseline has already begun to warm and flatten, with tree forms blurring into unstable vertical masses and the path losing crisp topological continuity. By the late frames, the difference becomes clear. The coherence-first rollout remains legible as a wooded path with near-field occlusion and background separation; the baseline loses scene topology through geometric distortion, brightness drift, and fragmented silhouettes.

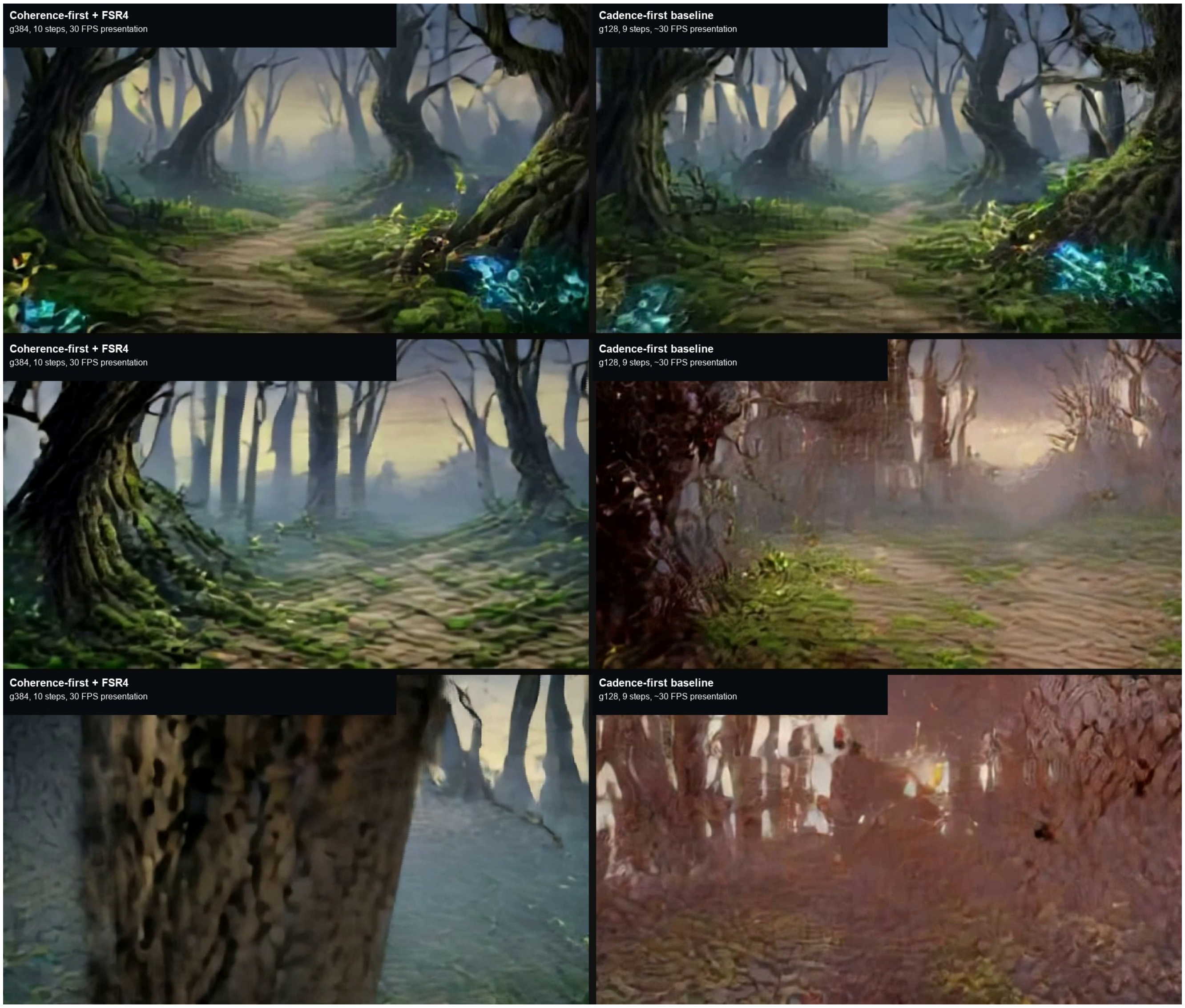

Figure 3. Forest rollout under the matched-route, compute-normalized comparison. Left: coherence-first branch (g384, 10 steps) reconstructed to a 30 FPS presentation stream with model-guided FSR4 frame generation. Right: cadence-first baseline (g128, 9 steps) generated natively at about 30 FPS presentation cadence. Both sequences begin from the same seed and follow the same deterministic route. The early frames are similar, but the baseline loses path continuity, tree identity, and depth / fog layering earlier, while the coherence-first branch remains readable deeper into the rollout.

### 7.2 Sword / held item: stability with a persistent first-person anchor

The sword scene provides the strongest secondary evidence because the held item adds a persistent first-person anchor. Here again the two branches begin close to one another. As the rollout proceeds, the coherence-first branch maintains both the sword silhouette and the larger environmental structure - the path, rock groupings, and hillside layout - more consistently. The cadence-first branch retains a sword-like object longer than it retains a coherent landscape, but the surrounding world loses stable terrain structure and coherent object boundaries earlier. The improvement is therefore not just background detail; it is viewpoint-consistent world persistence.

At the same time, the sword scene should be read as a special case rather than a generic environment. The held item is intended to remain nearly fixed in screen space while the background responds to camera motion. That makes the scene useful for judging first-person persistence, but it can also make motion-sensitive diagnostics and operating-point changes less representative than forest, desert, or snow.

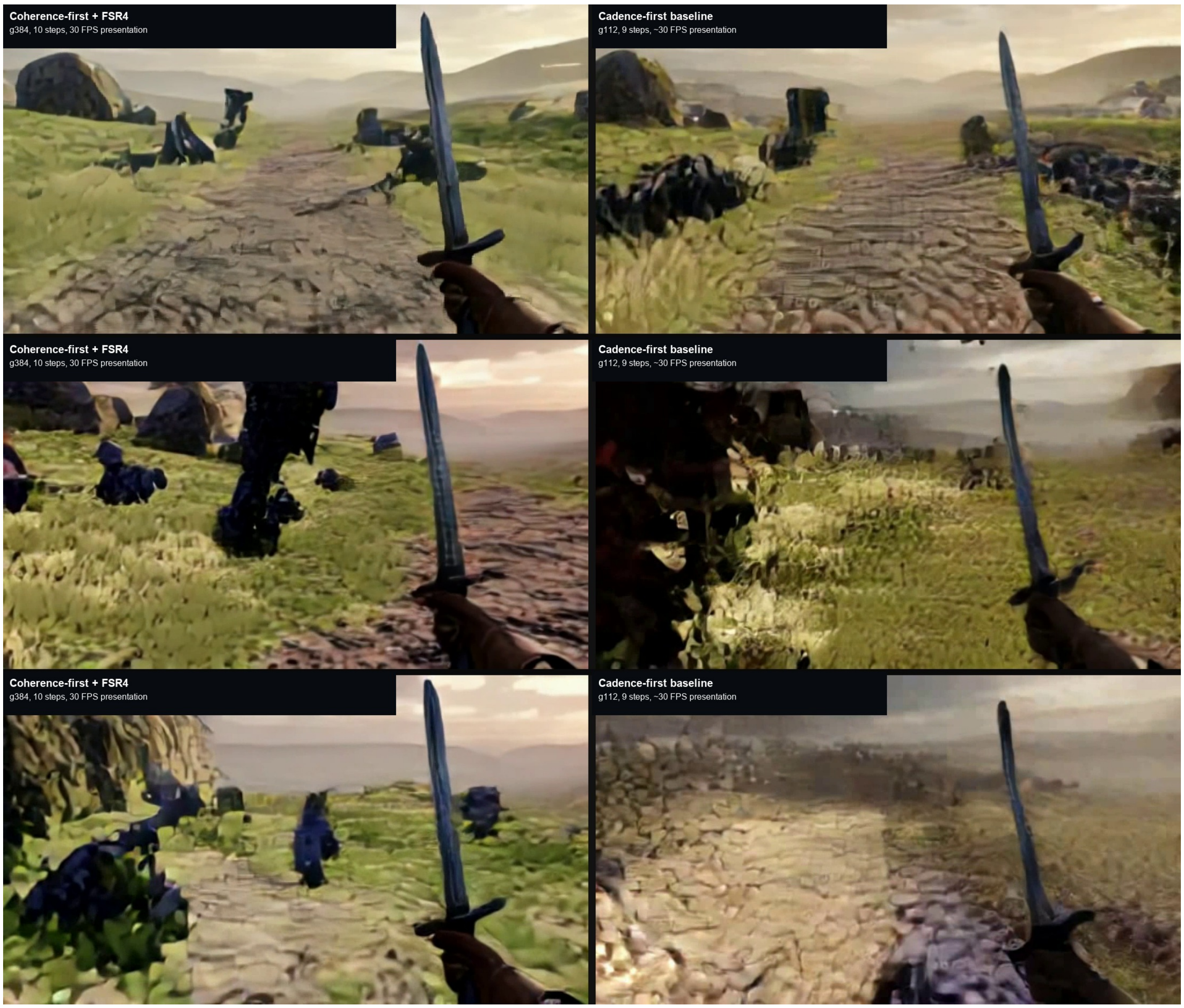

Figure 4. Held-item first-person comparison at 2.5 s, 9.0 s, and 11.0 s. The visible sword provides an additional viewpoint anchor, making temporal stability easy to judge. The coherence-first branch preserves both the sword silhouette and the surrounding landscape consistently better than the cadence-first baseline, whose environment degrades into unstable terrain and smeared object boundaries.

### 7.3 Desert and snow: cross-scene support

The desert scene supports the main claim in a sparser environment. Because there is less clutter, failures in large-scale geometry are especially visible. The coherence-first branch preserves the road, cactus identities, and distant mesa silhouettes longer, whereas the baseline increasingly blurs object boundaries and loses stable horizon structure.

The snow scene is harder for both methods, likely because bright, low-texture regions make the scene more sensitive to instability. Even so, the coherence-first branch still retains clearer track boundaries, tree lines, and spatial layering than the baseline, which develops blocky reflective artifacts sooner. For that reason, snow works best as supplementary evidence rather than as the lead figure.

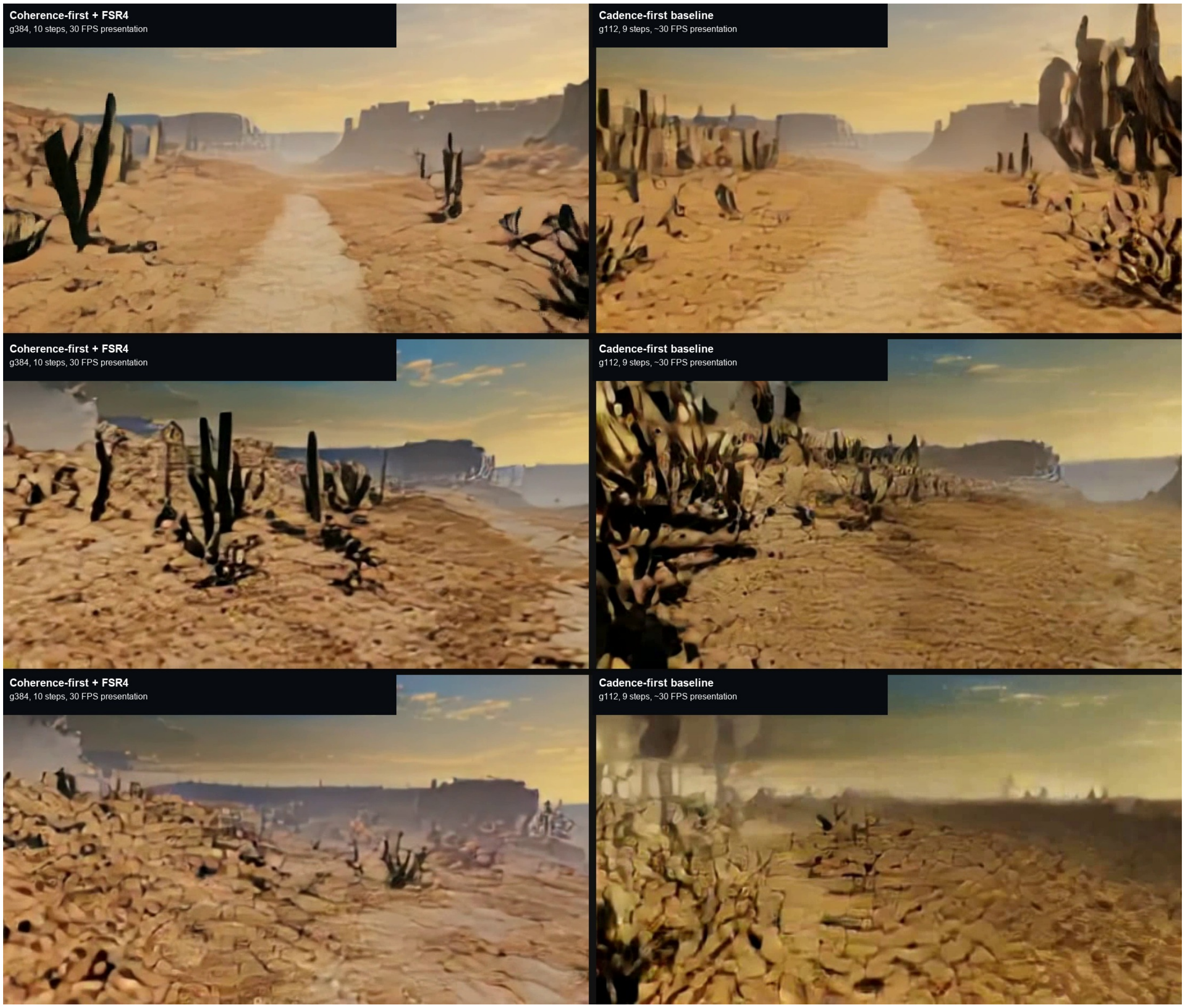

Figure 5. Cross-scene check in a sparse desert environment. The coherence-first branch maintains road structure, cactus identity, and large distant mesa silhouettes longer than the cadence-first baseline. Because the scene is sparse, failures in large-scale geometry are especially easy to see.

**Exploratory operating-point probe**

- The main paper comparison is coherence-first g384 / 10-step versus the cadence-first baseline on matched deterministic rollouts.
- The g512 / 12-step sword run used a more lenient timescale and is a separate operating-point probe. It is not part of the main fixed-route benchmark and is not used to support the compute-normalized headline comparison.

### 7.4 Exploratory heavier-run probe

This subsection is intentionally outside the main comparison. The paper's primary evidence is the fixed-route comparison between the selected coherence-first branch and the cadence-first baseline. The `g512`, 12-step sword run is an exploratory heavier probe only; it is not included in the main fixed-route benchmark and should not be read as a third branch in the primary comparison.

Within that limited probe, quality was not monotonic along this tested operating-point change. The selected `g384`, 10-step setting outperformed the heavier `g512`, 12-step run. The heavier run still looked better than the cadence-first baseline, but it was visibly worse than the chosen operating point in path readability, scene cleanliness, and object stability. A plausible interpretation is that the sword scene places competing demands on the model: a near-stationary first-person foreground object must remain locked to the view while the background continues to move and reorganize. Extra context and steps may not help if they reinforce the foreground constraint while the surrounding world becomes less clean. This is only an interpretation of one probe, not a demonstrated mechanism. The safer conclusion is that the operating regime must be selected empirically rather than assumed to scale upward indefinitely.

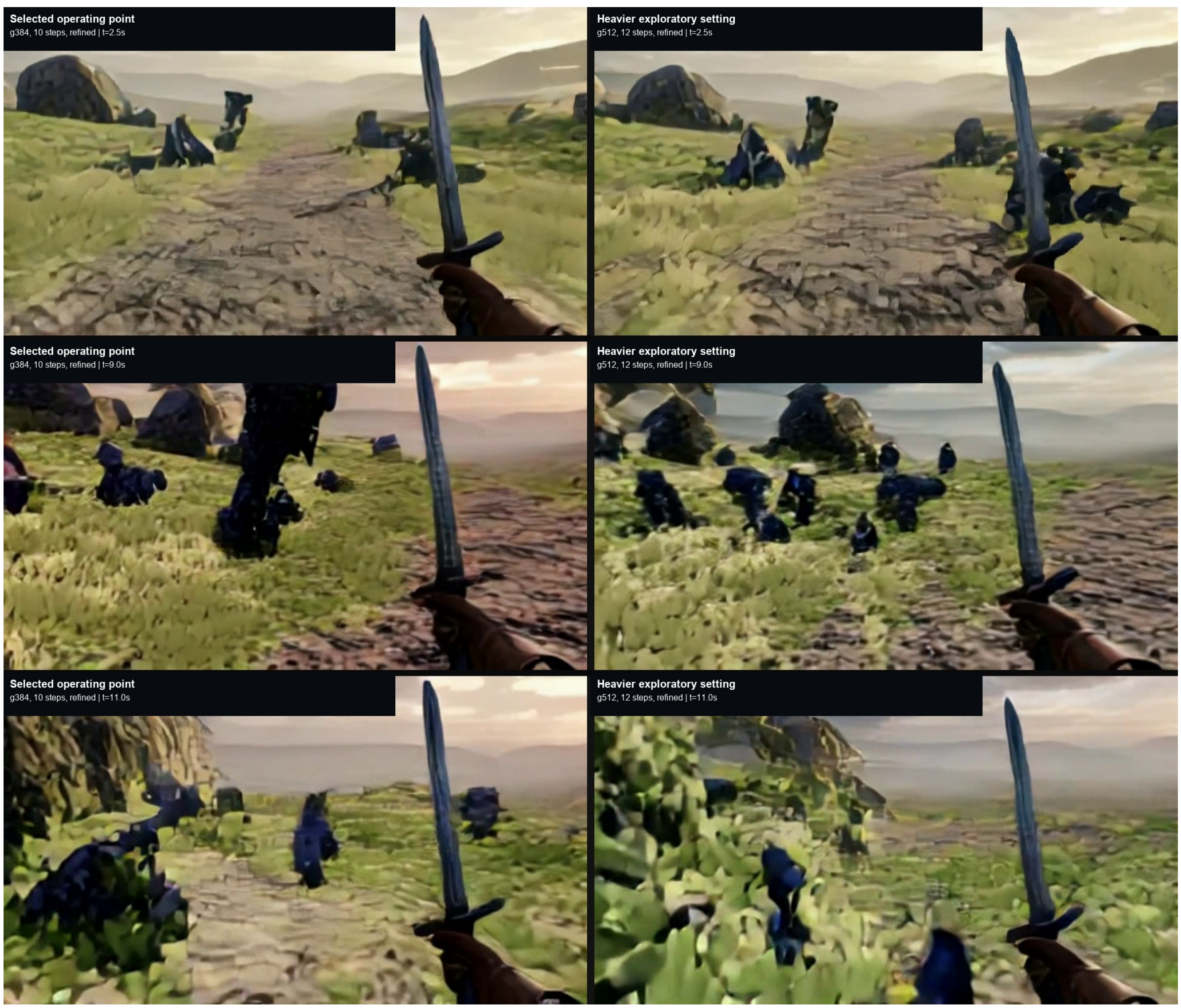

Figure 6. Exploratory operating-point probe for the sword scene, outside the main fixed-route comparison. Left: selected quality setting (g384, 10 steps, refined). Right: heavier setting (g512, 12 steps, refined). The heavier run remains better than the cadence-first baseline but is visibly worse than the selected operating point, showing local non-monotonicity along this tested context/step change.

## 8. Discussion

The evidence supports a specific systems conclusion. For this world-model setting, limited compute can be better spent on stronger anchor states than on higher native cadence. Once the base world state becomes unstable, simply producing more of it more often does not recover a believable environment. By contrast, if anchor states are coherent enough, reconstruction can preserve a smoother presentation without erasing the gain in world quality. The paired videos are informative because both branches often start near the same visual state and only separate clearly over time.

For the same reason, the contribution should not be reduced to generic post-processing. The decisive change is upstream: the coherence-first branch generates a more stable world state before reconstruction. The FSR4 bridge then reconstructs presentation from latent-delta motion and depth cues. In other words, the gain is not solely attributable to presentation-stage filtering; the reconstruction stage is leveraging internal scene dynamics rather than masking poor generation.

There is also a practical domain-alignment reason the bridge is promising. Waypoint is trained on large-scale game footage paired with controls [3], while AMD describes both FSR Upscaling and FSR Frame Generation as ML systems trained on high-quality game data [10,11]. This does not establish compatibility by itself, but it means the generator and the reconstructor both operate in a game-like distribution of motion, edges, and temporal structure. That is another reason this result is better described as a generator-reconstruction pairing than as generic post-processing.

### 8.1 On quantification

We include lightweight rollout-level temporal metrics rather than leaving the evidence purely qualitative. Two evaluation modes are used. Presentation-stream metrics compare adjacent frames on the final 30 FPS presentation videos. Anchor-aligned metrics sample both outputs at 15 Hz; for the coherence-first branch this corresponds to generated anchors, while for the baseline it is a 15 Hz subsample of its approximately 30 FPS presentation stream.

LPIPS is used as a temporal perceptual-change proxy because the paper's claim is about temporal drift and long-horizon scene coherence rather than exact pixel preservation. In all four scenes, the coherence-first branch has lower adjacent-frame LPIPS than the cadence-first baseline in both modes. Presentation-stream LPIPS means are 0.0449 vs 0.0677 for forest, 0.0404 vs 0.0568 for sword, 0.0365 vs 0.0562 for desert, and 0.0480 vs 0.0660 for snow. Anchor-aligned LPIPS means show the same trend: 0.0747 vs 0.0931, 0.0664 vs 0.0753, 0.0590 vs 0.0765, and 0.0763 vs 0.0959, respectively.

These metrics are supportive rather than definitive, since the scene set is small and curated and the main claim concerns long-horizon perceptual coherence.

SSIM was also computed as a secondary structural check. Presentation-stream SSIM favors the coherence-first branch in all scenes, while anchor-aligned SSIM is more mixed. One reason is that anchor-aligned comparison samples the coherence-first branch at generated anchors that are farther apart in presentation time, so lower SSIM can reflect genuine scene evolution rather than only instability. We therefore use both metrics to support the visual argument, not to replace it: the coherence-first branch stays stable longer, and the cadence-first branch diverges earlier in both the videos and the rollout-level LPIPS curves.

### Table 6. Temporal LPIPS summary across scenes

| Scene | Anchor LPIPS | Baseline anchor LPIPS | Full-stream LPIPS | Baseline full LPIPS |
|---|---|---|---|---|
| Forest | 0.0747 | 0.0931 | 0.0449 | 0.0677 |
| Sword | 0.0664 | 0.0753 | 0.0404 | 0.0568 |
| Desert | 0.0590 | 0.0765 | 0.0365 | 0.0562 |
| Snow | 0.0763 | 0.0959 | 0.0480 | 0.0660 |

### Table 7. Temporal SSIM summary across scenes

| Scene | Anchor SSIM | Baseline anchor SSIM | Full-stream SSIM | Baseline full SSIM |
|---|---|---|---|---|
| Forest | 0.5047 | 0.5404 | 0.7123 | 0.6118 |
| Sword | 0.5729 | 0.5779 | 0.7192 | 0.6014 |
| Desert | 0.6529 | 0.6386 | 0.7815 | 0.6833 |
| Snow | 0.6364 | 0.6324 | 0.7856 | 0.7283 |

### Table 8. Supplementary turn-onset response proxy across scenes

A simple control-to-visible-turn proxy suggests that the coherence-first branch can make turning visually detectable earlier in forest, desert, and snow, even though its true anchor cadence is lower. The detector measures horizontal motion energy in a central crop of the final presentation video after the scripted turn command begins, using a smoothed optical-flow signal and a sustained-threshold onset rule. We include this as supporting evidence rather than a formal latency claim. The sword scene is an outlier, plausibly because a near-stationary held object occupies part of the view while the background moves, so the detector is still scene-sensitive.

| Scene | Coherence-first onset | Cadence-first onset | Advantage |
|---|---|---|---|
| Forest | 0.467s | 1.767s | 1300 ms earlier |
| Desert | 0.533s | 2.100s | 1567 ms earlier |
| Snow | 0.133s | 1.567s | 1433 ms earlier |
| Sword | 3.767s | 2.300s | 1467 ms later |

## 9. Limitations

This study is deliberately narrow. First, it is a proof of concept on a Windows DirectML fallback path, not a claim about the best achievable Waypoint runtime or the final performance envelope of the model. Second, the strongest demonstrated result is 15 FPS latent anchors reconstructed to 30 FPS presentation. The paper does not show native 60 FPS generation, and it does not establish a finished real-time 60 FPS product pipeline. The original 10 -> 60 vision remains motivation rather than reported scope.

Third, compute-normalized here is an approximate same-GPU, same-timescale operating point, not a hardware-counter proof of exact FLOP parity. The submitted videos should be read on their presentation timeline: both main branches are slowed by the same factor on the local GPU, so the reported 15 FPS and 30 FPS values are presentation cadences rather than live interactive throughput. The saved manifests record rollout generation statistics, but they are not a complete independent timing study of every presentation-stage cost, so the timing claim should be read as operating-point accounting rather than a microbenchmark. Fourth, the evaluation is qualitative and limited to a small set of curated deterministic scenes. That is appropriate for an initial systems note on coherence, but it is not a substitute for broader perceptual studies, formal temporal-consistency metrics, or larger backend sweeps. Fifth, only one heavier operating-point probe is included in the curated asset set. That is enough to establish local non-monotonicity in at least one case, but not enough to map the full operating surface.

## 10. Conclusion

The main result of this work is straightforward but intentionally bounded: in the tested Windows-stack operating points, a coherence-first world-model pipeline - generate fewer, better anchor frames, then reconstruct presentation with model-guided FSR4 frame generation - produces a more stable and readable world than the lower-context cadence-first baseline under an approximately matched same-GPU, same-timescale budget. The evidence is strongest in the forest and sword scenes, supported by desert and still visible in snow. Separately, the exploratory heavier-run probe suggests that more context and more denoise steps do not automatically improve quality.

This does not yet amount to a production rendering stack. It is, however, a proof of concept for a practical rendering strategy in generative worlds. The clearest contribution is to reframe the optimization target. For this class of system, the relevant question is not only how many frames can be generated natively, but how best to spend limited compute so that the world remains coherent over time.

## Author note

Authorship note: this work was conceived, implemented, and evaluated as an independent project by Paweł Katarzyński. The listed university affiliation identifies the author's student enrollment and is not a claim of institutional sponsorship or funding. AI tools were used as engineering and drafting assistance, but experimental design, interpretation, and final authorship remain with the author.

# Supplementary figure included for completeness

## Figure S1. Snow comparison

The snow scene is harder for both methods than forest or sword, but it still supports the main claim. Both branches begin from a bright, low-texture winter path. The coherence-first branch preserves track boundaries, tree lines, and scene layering more effectively, while the baseline develops blocky reflective artifacts and unstable vertical structure earlier in the rollout.

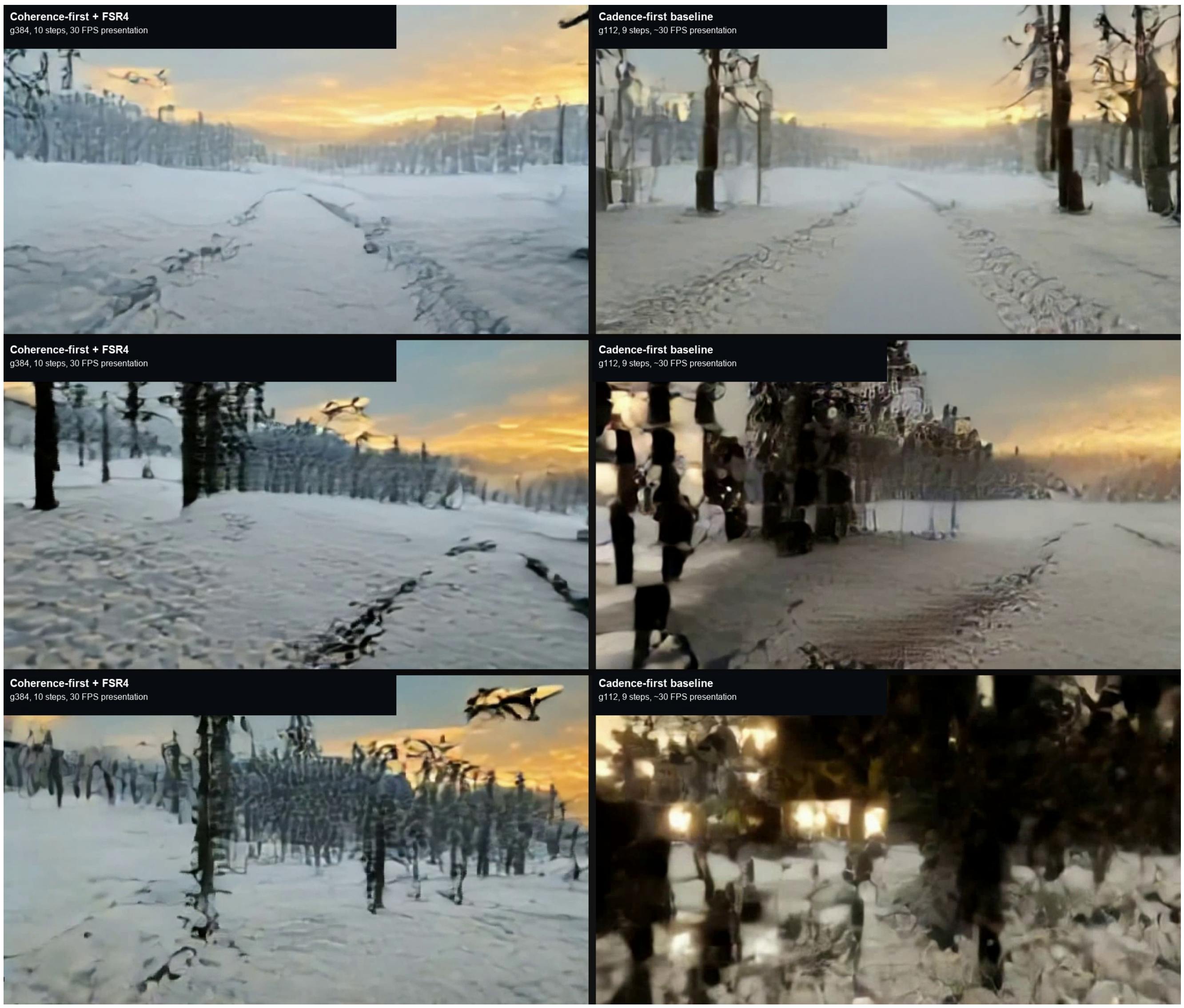

Figure S1. Snow-scene comparison in a bright, low-texture environment. Both branches are less stable here than in forest or sword, but the coherence-first branch still preserves track boundaries, tree lines, and scene layering more effectively than the cadence-first baseline.